# An Overview of Path Analysis: Mediation Analysis Concept in Structural Equation Modeling

Hashem Salarzadeh Jenatabadi
Department of Science and Technology Studies
University of Malaya, Kuala Lumpur, Malaysia
salarzadeh@um.edu.my
+60178777662

**Abstract**

This paper provides a tutorial discussion on path analysis structure with concept of structural equation modelling (SEM). The paper delivers an introduction to path analysis technique and explain to how to deal with analyzing the data with this kind of statistical methodology especially with a mediator in the research model. The intended audience is statisticians, mathematicians, or methodologists who either know about SEM or simple basic statistics especially in regression and linear/nonlinear modeling, and Ph.D. students in statistics, mathematics, management, psychology, and even computer science.

**Introduction**

Panel data, regression, and analysis of variance (ANOVA) are the most methodologies that have been using in so many studies [1-6]. In recent years, path analysis with SEM has attracted the attention of many researchers and organizations as a commonly adopted method used for data analysis tasks in various disciplines like computer science [7, 8], education [9, 10], engineering [11-15], and management [16-21].

The analyses of path and factors are both integrated and incorporated into SEM analysis forming a hybrid equation with both multiple factors for each specified variable, i.e., latent factors or variables, and paths joining the latent indicators together [22]. If the factors composite scores (or index items or composite) replace the unobserved (latent) variables and their indicators, and in case the observed (manifest) items are connected together through arrows, the resulting model is named as the path model. Therefore, it can be concluded that path analysis is a specific type of SEM method [22].

According to Garson [22] and Kline [23], SEM with a single indicator (observed variable) is also considered as a path analysis. Using the software for SEM as a model in which each indicator has multiple variables without any direct effects (arrows), attaching the indicators is considered as a kind of factor analysis. Nevertheless, Garson believes that using SEM software with each factor containing only one measurement indicator is also a sort of path analysis. In a path analysis the observed (manifest) variables are typically used to form a composite of sum scores of the factors or variables of each scale in order to gauge an unobserved (latent) construct [24].

Single-indicators (observed indicators) are graphically specified by squares and unobservable (latent) indicators represented by ovals. A variables model represented solely by squares is known as a path model. However, a model with variables indicated by squares instead of ovals attached to the variables through arrows is called a structural model.

Accordingly, the differences between path analysis and SEM analysis can be summarized as below. It should be highlighted that path analysis is considered to be a specific form of SEM analysis. SEM analysis utilizes unobserved (latent) indicators gauged by many observed indicators, while path analysis employs just observed measurement generated by the sum scores of the multiple factors, which are utilized to compute the unobserved (latent) constructs. Nevertheless, SEM and path analyses have a common feature that makes them similar, i.e., both are utilized to determine whether or not the overall model is fit to suit the gathered data and investigate the individual hypotheses.

The main trend of the current research was elaborated upon in the previous section in which indication of the path analysis, a particular sort of SEM technique, was attempted. This indication is also applicable for analysis of the hypothesized links within the model. AMOS 6.0 [25], a convenient graphical SEM software program, was used for this analysis. AMOS routinely creates equations for the model after a diagram is drafted on the computer.

**Construction of the Path Model for Estimation**

The above-mentioned phase is followed by another step, which concentrates on the formation and construction of a path model based on the outcomes resulting from the multiple regression analyses. In the path model "double-headed or single-headed arrows" and squares represent the structural relationships and their directions among the variables. The squares represent the observed (manifest) variables, while unobserved (latent) indicators are indicated by ovals (which in structural model were represented by squares). The single-headed arrows show the causalities or the structural relationships between the dependent (mediating) and independent variables, and double-headed arrows show the correlations that exist between the independent variables.

The variables that ensued from the regression analyses were also employed in the path model as endogenous (dependent), exogenous (independent), and mediating (intervening) variables. In a path model, the variables that do not have any obvious causes, i.e., there are no arrow signs directed to them, are considered as exogenous variables. However, if there is a correlation between the exogenous variables, a double-headed arrow is used to indicate the correlation between them. There are also variables that are indicated by single-headed arrows directed to them, which represent a regression (causal) relationship with an exogenous variable. These kinds of variables are endogenous in fact although it is possible to use mediating variables as both endogenous and exogenous ones. Therefore, they both have outgoing and incoming causal arrows in the graphical path diagram [22, 26].

Furthermore, this step of the path model deals with terms of disturbance or residual error for each observed endogenous variable. For the endogenous factor, the error term represents unexplained variance, i.e., the unmeasured items effects, and the resulting measurement error. The paths from endogenous variables, also known as regression paths, towards their term of disturbance are also indicated by single-headed arrows. These regression paths are shown by 1, which indicates their "initial values" allowing SEM (AMOS) to assess and evaluate the term of the disturbance variance [22, 27].

**Testing Model Fit**

The fit index statistic tests the consistency between the predicted and observed data matrix by the equation [28]. One of the differences that exist between the SEM technique and regression method is that the former one does not have any single statistical test applicable for evaluation of model predictions "strength" [29]. In this regard, Kline [23] believed that there are "dozens of fit indexes described in SEM literature, more than any single model-fitting program reports". However, according to Hair, Black [29] and Garson [22], the chi-square fit index, also known as chi-square discrepancy test, is considered as the most fundamental and common overall fit measure. Thus, in a good model fit the value of chi-square should not be very significant, i.e., p>0.05 [29]. However, one problem usually experienced through this test relates to the rejection probability of the model having direct interaction with the sample size. Moreover, the sensitivity level of chi-square fit index is very high, especially, towards the multivariate normality assumption violations [22].

Many indexes have been introduced and developed to avert or reduce the problems related to the chi-square fit index. Some of the indexes included in the absolute fit indexes are as follows:

   a) *"Normal Chi-Square Fit Index" (CMIN/DF):*

Normal chi-square fit index, $\chi^2/df$, serves to adjust the testing of chi-square according to the sample size [27]. A number of researchers take 5 as an adequate fit value, while more conservative researchers believe that chi-square values larger than 2 or 3 are not acceptable [22].

   b) *"Goodness-of-Fit Index"*[30]:

GFI is utilized for gauging the discrepancy level between the estimated or predicted covariances and resulted or observed ones [31].

$$GFI = 1 - \left[\frac{\max[(\chi^2 - df)/n, 0]}{\max[(\chi^2_{null} - df_{null})/n, 0]}\right]$$

The allowable range for GFI is between 0 and 1, where 1 indicates a perfect fit, which demonstrates that measures equal to or larger than 0.90 signify a 'good' fit [22].

   a) *Adjusted Goodness-of-Fit Index"(AGFI)* [32]:

AGFI is utilized for adjustment of the GFI relating the complexity of the model.

$$AGFI = 1 - \left[(1 - GFI)\frac{d_{null}}{d}\right]$$

The measuring of AGFI is between 0 and 1, in which 1 or over 1 (AGFI>1.0) signifies a perfect fit, nevertheless, it cannot be bounded below 0, i.e., (AGFI<0). As in the case of GFI, AGFI values equal to or bigger than 0.90 signify a 'good' fit [22].

b) *"Root Mean Square Residual" (RMR)*:

RMR shows the mean squared amount's square root, which distinguishes the sample variances and covariances from the corresponding predicted variances and covariances [33]. The assessment relies on an assumption that considers the model to be correct. The smaller the RMR, the more optimal the fit is [22].

c) *"Root Mean Square Error of Approximation"* (RMSEA) [34]:

RMSEA is employed to gauge the approximation error in the population.

$$RMSEA = \left[\frac{(\chi^2 - df)}{(n-1)df}\right]^{1/2}$$

In cases where the RMSEA value is small, the approximation is believed to be optimal. An approximately 0.05 or smaller value of RMSEA means a more appropriate and closer model fit in connection with the degrees of freedom. Nevertheless, between 0.05 and 0.08 displays the most preferable status and the more optimal fit results [35].

In addition, the following indexes are also included in the incremental fit measures:

a) *"Normed Fit Index or Bentler Bonett Index"* (NFI):

Normed Fit Index or Bentler Bonett Index or NFI is applicable to contrast and compare the fit of a suggested model against a null model [36].

$$NFI = \left[\chi^2/df_{(Null\ Model)} / \chi^2/df_{(Proposed\ Model)}\right] / \left[\chi^2/df_{(Null\ Model)}\right]$$

This index defines all the observed variables as uncorrelated. The values of NFI range between 0 and 1, where 0.90 signifies an optimal fit [22].

a) *"Tucker Lewis Index or Non-Normed Fit Index"* (TLI or NNFI):

The TLI or NNFI index is used to gauge parsimony, which is applicable through the evaluation and assessment of the degrees of freedom of the suggested model to the degrees of freedom of the null model [36].

$$NFI = \left[\chi^2/df_{(Null\ Model)} / \chi^2/df_{(Proposed\ Model)}\right] / \left[\chi^2/df_{(Null\ Model)} - 1\right]$$

However, it is not certain whether TLI can vary from 0 to 1. A fit of model is required to possess a TLI that is larger than 0.90 [36, 37].

b) *"Comparative Fit Index"* (CFI) [38]:

CFI is not only less affected by the sample size, but also based on comparison of the hypothesized model to the null model [23].

$$CFI = 1 - \left[\frac{\max[(\chi^2 - df), 0]}{\max[(\chi^2 - df), (\chi^2_{null} - df_{null}), 0]}\right]$$

The values of CFI range between 0 and 1. However, its values need to be a minimum of 0.90 to be usable for a model fit [22].

**Mediation Model**

Most studies focus on the relations that exist between two variables, X and Y, which have been generously dealt with in various writings concerning the conditions under which Y is possibly affected or caused by X. Randomizing the units to X value and units independence within and across X value are also contained in these conditions.

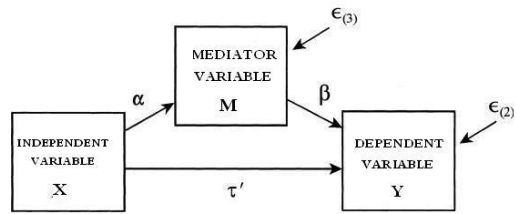

The mediation model seeks to discover and explicate the underlying mechanism of an observed relationship existing between a dependent and an independent variable through including a third explanatory variable, which is normally known as a the mediator variable. However, the hypothesis of a meditational model is not related to a direct causal relationship between the dependent and independent variable, but the hypothesis assumes that the independent variable as the main cause of the mediator variable, which, consequently, results in the dependent variable. Therefore, it can be claimed that the mediator variable seeks to explain the nature of the relationship between the dependent variable and the independent variable [39].

Direct effect=τ´

Indirect effect=αβ

Total effect= αβ + τ´

The above figure displays a simple Mediation model. The simplest Mediation model indicates the addition of a third variable to the independent variable and dependent variable relationship, which enables the independent variable (X) to cause the mediator (M), and the resulting mediator variable (M) to cause the dependent variable (Y), namely:

*Independent variable → Mediator variable → Dependent variable*

It should be noted that the relationship between X and Y is via the direct and mediated effect indirectly causing X to affect Y through M. The mediation model can be dichotomized into two more models: theoretical model, corresponding to unobservable relationship among indicators, and empirical model, corresponding to statistical analysis of actual data [40]. The relevant study tries to infer the true state of mediation from observations. However, some qualifications are attributable to this simple dichotomy, which is, generally, interested in justification of a research program to conclude that a third variable is mediating in the relationship.

**Mediation Regression Equations**

There are three main approaches that are commonly employed for analysis of the statistical mediation model. These approaches are: 1) causal (first) step; 2) difference in coefficients (second step); and 3) product of coefficients (third step). The required data used in these three approaches is mainly obtained from the three regression equations, displayed below:

$$Y = \alpha_1 + \beta_1 X + \varepsilon_1 \qquad (1)$$

$$Y = \alpha_2 + \beta_2 X + \beta_M M + \varepsilon_2 \qquad (2)$$

$$M = \alpha_3 + \beta_3 X + \varepsilon_3, \qquad (3)$$

In the above equations, Y is considered as the dependent variable; $\alpha_1$, $\alpha_2$ and $\alpha_3$ are intercepts; and M indicates the mediator; X represents the independent variable; $\beta_1$ indicates the coefficient related to the dependent and independent variables; $\beta_2$ shows the coefficient connecting the dependent variable to the independent one, and, ultimately, adjusting them for the mediator; $\beta_M$ represents the coefficient linking the mediator indicator to the dependent variable adjusted for the independent one; $\beta_3$ indicates the coefficient connecting the independent to the mediator variable; and, finally, $\varepsilon_1$, $\varepsilon_2$, and $\varepsilon_3$ indicate the residual terms. Nevertheless, it is noteworthy to mention that the mediation functions can be modified to produce both nonlinear and linear effects as well as M and X interactions in Equation (2).

The most common approach employed for the assessment and evaluation of the Mediation model is the first or causal steps approach. The causal steps approach has been delineated in the works of some researchers, such as Baron and Kenny [41]; Kenny, Kashy [42]; Judd and Kenny [43]; and Judd and Kenny [44]. For establishment of the mediation model, the Baron and Kenny approach suggests four steps, namely, in the first step, a strong relation between the dependent and independent variables is required for Equation (1). In the second step, Equation (3) requires a significant relationship between the hypothesized mediator and the independent indicator. Next, a significant mediator variable is required to be related to the dependent variable. However, both mediating and independent variables are predicting the dependent variable in Equation (2). Finally, in the fourth step, the coefficient connecting the dependent variable to the independent one is required, which needs to be greater (in absolute value) than the coefficient connecting the dependent variable to the independent one in the regression analysis in which both the mediating and independent variables, in the unique equation, are predictors of the dependent variable.

The causal steps approach, mentioned above, is the most common method utilized for assessment of the mediation model. However, this approach has a number of limitations, which are elaborated upon in this part. In the single-mediator model, the mediation effect can be computed in two ways, namely, $\widehat{\beta_3}\widehat{\beta_M}$ or $\widehat{\beta_1} - \widehat{\beta_2}$ [45]. The indirect or mediated effect value, calculated through the coefficient difference, $\widehat{\beta_1} - \widehat{\beta_2}$, in Equations (1) and (2), adjusts with a decrease of the independent factor effect on the dependent factor while corresponding to the mediation factor.

The product of coefficients are generated from the mediated or indirect effect, which involves assessment of the product of $\widehat{\beta_3}$ and $\widehat{\beta_M}$, $\widehat{\beta_3}\widehat{\beta_M}$ and estimation of Equations (2) and (3) [46]. This is due to the fact that mediation depends on the modification extent made in the mediator, $\widehat{\beta_3}$, by the program as well as the extent of the effect of the mediator on the produced variable, $\widehat{\beta_M}$. Next, the significance is checked through dividing the result by the standard error of the ratio, which is compared and contrasted to a standard normal distribution.

MacKinnon, Warsi [47] presented the algebraic equivalence of the $\widehat{\beta_1}-\widehat{\beta_2}$, and $\widehat{\beta_3}\widehat{\beta_M}$ measures of the mediation for normal theory OLS and MLE of the mediation regression models. Concerning the multilevel modelling [48], probit or logistic regression modelling [45], and analysis with survival data [49], the estimators of the mediated effect, $\widehat{\beta_3}\widehat{\beta_M}$ and $\widehat{\beta_1}-\widehat{\beta_2}$, are not always equivalent, and the two similar yields need to undergo some transformation [45].

**Standard Error of the Mediated Effect**

The multivariate delta method can be used as a common formula to find the standard error of the mediated effect [50, 51]. The indirect effect asymptotic standard error can be obtained through Equation (3.4) below [52]:

$$\sigma_{\widehat{\beta_3}\widehat{\beta_M}} = \sqrt{\sigma^2_{\widehat{\beta_3}}\hat{\beta}^2_3 + \sigma^2_{\widehat{\beta_M}}\hat{\beta}^2_3} \qquad (4)$$

Another formula that can be utilized to obtain the standard error of $\widehat{\beta_1}-\widehat{\beta_2}$ and $\widehat{\beta_3}\widehat{\beta_M}$, has been elaborated and delineated by MacKinnon, Lockwood [53]. However, the research that is based on simulation shows that the standard error of estimator in Equation (4) reveals that the sample size low bias should be a minimum of 50 in models of single-mediation [47]. In case a model's mediator number is more than one, the standard error of at least 100-200 sample size is accurate [54]. The resulting outcomes with similar features can be applied to positive and negative path values standard errors as well, while larger models contain multiple dependent, independent, and mediating indicators [55].

**Confidence Limits for the Mediated Effect**

The standard error of $\widehat{\beta_3}\widehat{\beta_M}$ is also applicable for examining the statistical significance of it as well as constructing confidence for the mediated effect restrictions, as shown in Equation (5) below:

$$\widehat{\beta_3}\widehat{\beta_M} \pm z_{1-w/2} * \sigma_{\widehat{\beta_3}\widehat{\beta_M}} \quad (5)$$

Some scholars who support bootstrap analysis and simulation studies of the mediated effect reveal [56] that confidence limits based on the mediated effect normal distribution [47] can hardly be precise and errorless. The confidence intervals of the mediating effect strongly lean to move towards the left side of the true value of the mediating effect for mediating effects that are positive. They also have a strong tendency towards the right of the negative mediating effects [54, 57]. The limits of asymmetric confidence based on the estimation of bootstrap and product distribution can contribute to the process in a more effective fashion than the afore mentioned tests [55].

**Distribution of the Product**

The outcome produced by two random variables normally distributed will be normal distribution in particular cases alone [58, 59]. This clarifies and makes clear the inaccuracy of assessing the statistical significance techniques of the normal distribution based mediation. As an example we can refer to two standard normal random variables with a zero mean, for which the excess kurtosis equals six [60] in comparison with an excess kurtosis of zero mean for a normal distribution. An experiment by MacKinnon, Lockwood [55] revealed that compared to common methods, the results of significance tests done for the mediated effect according to the product distribution contained more accurate statistical power and especially type-I error rates.

**Assumption of the Single-Mediator Model**

According to MacKinnon, Fairchild [61], there are several significant assumptions that can be used for mediation tests. In the case of the effect of the estimator mediated by $\widehat{\beta_3}\widehat{\beta_M}$, the model supposes the residuals that are in Equations (2) and (3) are independent, while in Equation (2), the residual and M are considered as statistically independent [62]. The presence of XM interaction in Equation (3) is to be tested for approval; nevertheless, such an interaction is assumed to be in the Equation. If model assumptions are correctly specified, there may not be causal order misspecification, such as $Y \to M \to X$ instead of $X \to M \to Y$. Causal direction misspecification like mutual causation between the dependent variable and the mediator, misspecification ensued from unmeasured variables, which prompts factors in the mediation study, or misspecification resulting from inaccurate and imperfect and inaccurate measurement [63, 64]. As a result of the impossibility or improbability of carrying out testing these assumptions in most conditions and situations, the approval of the mediation relation does not appear to be possible [61].

**Complete Versus Partial Mediation**

An important task of a relevant researcher is to test to find out whether the mediation is complete or partial. This task is normally done to see if the $\widehat{\beta_2}$ is significant. In reality, this testing is to identify whether the relationship between the dependent and independent variables is comprehensively explicable of a mediator [65]. If the $\widehat{\beta_2}$ and the mediation are statistically significant, a partial mediator indicator may exist [61].

**Inconsistent and Consistent Models**

Inconsistent mediation models are those in which there is at least one mediating effect with a different sign in comparison to other direct or mediated effects in the same model [66, 67]. The relation of X to Y should be significant for interpretation of the outcome; however, there may be other cases in which the overall relation of X to Y is not significant despite the existence of mediation. McFatter [68] explained the widgets hypothetical pattern, which makes labourers. In this example X indicates intelligence, M represents the rate of boredom, and Y shows production of the widget. Intelligent labourers are likely to be bored, which, ultimately, leads to a reduction of their production rate. Nevertheless, workers who enjoy higher intelligence are more likely to produce more widgets. As a consequence, it is possible to actually have zero level of the overall relation between the number of the produced widgets and intelligence. However, it is possible that two opposing meditational processes exist concurrently in a model. Several other examples like the one presented above, provide sufficient demonstrations of such inconsistent effects [69, 70]. Nonetheless, inconsistent mediation is applied more commonly in multiple mediator models, in which various mediated effects have different symptoms. Inconsistent mediator effects are possible to be specifically critical in assessment and evaluation of counterproductive effects of tests, the manipulation of which can lead to mediated effects with opposing features [61].